\documentclass[lettersize,journal]{style}
\usepackage{amsmath,amsfonts}
\usepackage{algorithmic}
\usepackage{algorithm}
\usepackage{array}
\usepackage[caption=false,font=normalsize,labelfont=sf,textfont=sf]{subfig}
\usepackage{textcomp}
\usepackage{stfloats}
\usepackage{url}
\usepackage{verbatim}
\usepackage{graphicx}
\usepackage[%
    colorlinks=true,
    pdfborder={0 0 0},
    linkcolor=black,
    citecolor=black,
    urlcolor=blue,
]{hyperref}
\usepackage[style=ieee, maxnames=6, minnames=1, url=false]{biblatex}
\usepackage[normalem]{ulem} 
\usepackage{tikz}

\newcommand\submittedtext{%
  \footnotesize This work has been submitted to the IEEE for possible publication. Copyright may be transferred without notice, after which this version may no longer be accessible.}

\newcommand\submittednotice{%
\begin{tikzpicture}[remember picture,overlay]
\node[anchor=south,yshift=10pt] at (current page.south) {\fbox{\parbox{\dimexpr0.65\textwidth-\fboxsep-\fboxrule\relax}{\submittedtext}}};
\end{tikzpicture}%
}

\begin{document}

\title{Cross-Polarization Reduction in Kinetic Inductance\\ Detectors Based on Quasi-Lumped Resonators}

\author{Victor Rollano, Martino Calvo, Alejandro Pascual Laguna, David Rodriguez, Maria Teresa Magaz, Beatriz Aja,\\ Luisa de la Fuente, Daniel Granados, Alessandro Monfardini, Alicia Gomez
\thanks{Manuscript received July 7, 2025.}
\thanks{This work has received support from grants PID2022-137779OB-C41, PID2022-137779OB-C42, PID2022-137779OB-C43, JDC2023-051842-I and the research network RED2022-134839-T funded by the Spanish MCIN/AEI/10.13039/501100011033, by the EU “NextGenerationEU”/PRTR and by the “ERDF A way of making Europe”. IMDEA Nanoscience acknowledges financial support from the “Severo Ochoa” Programme for Centres of Excellence in R\&D (CEX2020-001039-S) and CAB from the the CSIC Research Platform PTI-001 and from “Tecnologías avanzadas para la exploración del Universo y sus componentes” (PR47/21 TAU-CM) project funded by Comunidad de Madrid, by “NextGenerationEU”/PRTR.}
\thanks{This article is an expanded version of the IEEE MTTS International Microwave Symposium (IMS 2025). (Corresponding author: Alicia Gomez)}
\thanks{V. Rollano, A. Pascual Laguna, D. Rodríguez, M.T. Magaz and A. Gomez are with Centro de Astrobiología (CSIC-INTA), Torrejón de Ardoz, E-28850 Madrid, Spain (email: vrollano@cab.inta-csic.es, agomez@cab.inta-csic.es).}
\thanks{M. Calvo and A. Monfardini are with the Institut Néel, CNRS and Université Grenoble Alpes, Grenoble 38042, France}
\thanks{B. Aja, and L. de la Fuente, are with Dpt. Ing. Comunicaciones, Universidad de Cantabria, 39005 Santander, Spain.}
\thanks{D. Granados is with IMDEA-Nanociencia, Cantoblanco, E-28049 Madrid, Spain.}}


\maketitle
\submittednotice

\begin{abstract}
Kinetic Inductance Detectors (KIDs) have emerged as a leading technology for millimeter- and submillimeter-wave astronomy due to their high sensitivity, natural multiplexing capabilities and scalable fabrication. In polarization-sensitive applications—such as Cosmic Microwave Background (CMB) studies—cross-polarization, or unintended response to the orthogonal polarization, poses a significant limitation to measurement fidelity. This work investigates the origin of cross-polarization in meandered Lumped Element KIDs (LEKIDs), with particular emphasis on the role of parasitic currents in the interdigitated capacitor. A comparative study between conventional LEKIDs and a quasi-lumped resonator design is presented, demonstrating that removing the capacitive element may improve cross-polarization discrimination, confirming the capacitor’s contribution to polarization leakage.
\end{abstract}


\section{Introduction}
\IEEEPARstart{K}{inetic} Inductance Detectors (KIDs) are superconducting detectors that exploit the Cooper pair breaking to sense when photons are absorbed \cite{zmuidzinas2012}. The changes in the superconducting carrier density due to photon absorption lead to an increase in the kinetic inductance ($L_k$) of the material. KIDs are typically implemented as planar microwave resonators, and the frequency shift induced by kinetic inductance change serves as a direct measurement of the incoming radiation. These superconducting resonators can be designed with slightly different resonance frequencies and coupled in parallel to a common transmission line. Hence, KIDs are intrinsically frequency-multiplexable, enabling the deployment of large-format detector arrays with minimal readout complexity \cite{day2003}. Their multiplexing capability, high sensitivity and technological-readiness, make KIDs ideal for astronomical instrumentation.

At (sub-)millimeter wavelengths, KIDs have been successfully integrated into a number of astronomical instruments, e.g.: the cameras NIKA\cite{catalano2014performance}, NIKA2 \cite{perotto2020calibration}, CONCERTO \cite{desert2025continuum}, AMKID \cite{reyes2025early}, MUSIC \cite{golwala2012music}, TolTEC \cite{wilson2020toltec}, MAKO \cite{swenson2012mako}; and the spectrometers SuperSpec \cite{shirokoff2014design}, u-Spec \cite{cataldo2019uspec}, DESHIMA \cite{endo2019wideband}. These instruments have demonstrated state-of-the-art sensitivity and spatial resolution, largely attributed to the scalability and performance of large KID arrays. More recently, KIDs have been selected for the PRIMA space telescope of the Probe-Explorers class mission of NASA, confirming their status as a leading detector technology for space-based applications \cite{Glenn2024prima}.

KIDs offer significant advantages in the millimeter-wave regime, where traditional coherent receivers approach the quantum noise limit. As incoherent detectors, KIDs have the potential to achieve sensitivities beyond this limit \cite{Zmuidzinas2003thermalnoise}, making them particularly attractive for next-generation astrophysical observations. Their applicability is especially relevant in experiments targeting the Cosmic Microwave Background (CMB), where high sensitivity and low-noise performance are critical for detecting faint polarization signals. Additionally, KIDs are strong candidates for dark matter search and other fundamental physics experiments \cite{aja2022canfranc}.

Superconducting resonators designs fall into two categories: distributed resonators, where the resonant mode is defined by the length of the superconducting line; and Lumped Element Resonators (LERs), where distinct inductive and capacitive components are combined in a compact layout to define the resonance frequency. Relating to the optical coupling mechanism, KIDs can be categorized as: antenna-coupled KIDs, where an antenna is used as a receiver and placed at the maximum current of a distributed resonator to maximize its response; and Lumped Element KIDs (LEKIDs), based on LERs, where a meandered line is employed as absorbing element. In this case, the current is uniformly distributed along the meander length, yielding to a uniform response along it.  
In addition to high sensitivity, polarization selectivity is often also required. Antenna-coupled KIDs have demonstrated excellent polarization purity, as the antenna—being the receiver—naturally provides good cross-polarization discrimination \cite{yates2025ultrasensitive, ferrari2018antenna}. In contrast, the polarization response in LEKIDs is primarily determined by the geometry of the inductor. Traditional meandered inductors are inherently polarization-sensitive, as they preferentially absorb incoming radiation with the electric field aligned parallel to the long segments of the inductor. In this case, to optimize polarization discrimination, the inductor design should therefore minimize absorption in the segments oriented perpendicular to the desired polarization direction, as have been proposed in recent studies \cite{mccarrick2018design, deory2023}. Contrarily, when seeking an equal response for both polarizations, Hilbert-like or double-meandered geometries can be employed \cite{calvo2013hilbert,dabironezare2025lens}. 

In any case, the potential contribution of the interdigitated capacitor to the polarization response remains an open question. Parasitic currents arising in the capacitor fingers \cite{eichler2017electron} can introduce cross-polarized response if optical absorption occurs there. This could degrade the polarization selectivity of the detector by increasing its cross-polarization response. Indeed, such effects may explain the elevated cross-polarization observed in previous studies \cite{deory2025polarimeter}, and they could become especially significant in the development of filled-array LEKID cameras where no coupling elements (e.g., horn or lens) are used. Different strategies can be employed, such as the fabrication of the capacitor in a higher superconducting-gap material (e.g., niobium), or the use of parallel-plate capacitors \cite{beldi2019high}. The use of focusing elements on the inductor, such as lenses or horns, can also reduce the contribution of the capacitor to the cross-polarization response \cite{dabironezare2025lens}. However, these strategies inevitably increase the complexity of developing plain LEKID cameras. In this context, the objective of this work is to investigate the role of the capacitor structure as a contributing factor to cross-polarization in meandered LEKIDs, with particular interest in advancing the development of an on-chip polarimeter \cite{deory2025polarimeter}.

\section{Design overview}
\begin{figure}[!t]
\centering
\includegraphics[width=0.9\columnwidth]{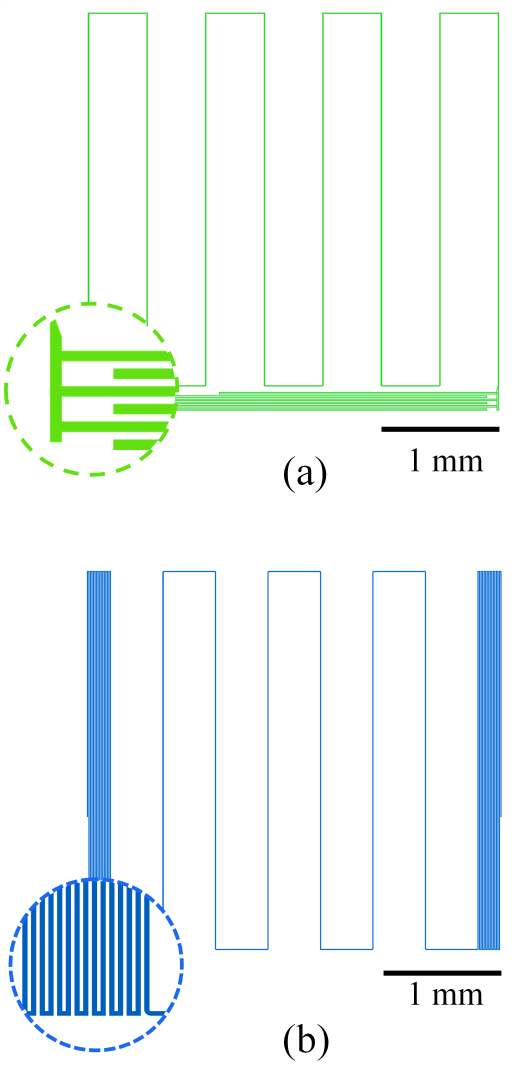}
\caption{(a) Design used for the LER-type detectors. The inset shows a zoomed region of the interdigitated capacitor. (b) Design used for the qLER-type detectors. The inset shows a zoomed region of the ‘compressed’ part of the inductor. A black scale bar indicating 1 mm is included.}
\label{fig_1}
\end{figure}

Two types of detectors have been designed: one based on a lumped-element resonator (LER) that incorporates an interdigitated capacitor, and another based on a distributed resonator. Fig. \ref{fig_1} illustrates the two designs, both having the same meandered absorber geometry to ensure comparable absorption characteristics for the intended polarization. 

The active region of both resonators, designed for W-band absorption, presents strips of width $2a = 3$ $\mu$m, length of $l = 3.086$ mm and space between them of $g = 440$ $\mu$m. For meandered absorbers, impedance matching to free space can be expressed in terms of the strip geometry ($a$,$g$) and the sheet resistance of the superconducting film prior to the superconducting transition ($R_s$). The procedure for determining the meander geometry is described in \cite{aja2020analysis}. The strip grating is matched to $\eta_0$ = 377 Ohm with a backshort at the rear side of a 0.275$\lambda$ thick Silicon substrate with dielectric constant $\epsilon_r$ = 11.9, which presents a capacitance effect. On the other hand, the strip grating at readout frequencies provides an inductive effect, which is used to design a resonator, with an interdigital capacitor for the LER-type detectors or extending it up to a $\lambda$/2 transmission line for the qLER detectors.

By orienting the fingers of the capacitor perpendicular to the long sections of the meander, it is possible to assess the extent to which the interdigitated capacitor contributes to the cross-polarization response.

\begin{figure}[!h]
\centering
\includegraphics[width=0.9\columnwidth]{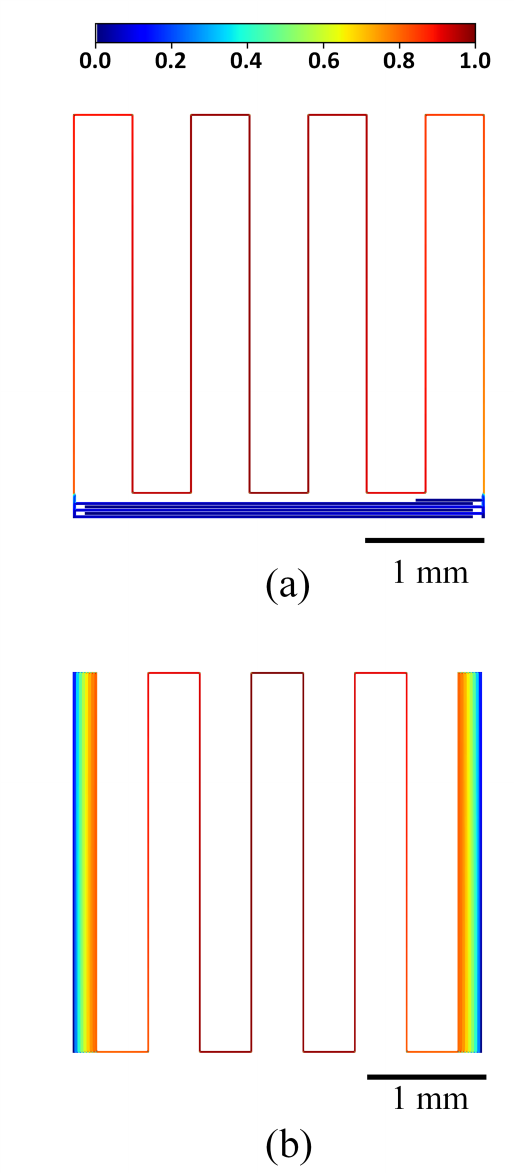}
\caption{Simulated current distribution for a (a) LER design and (b) qLER design. In both simulations the current flows uniformly across the active region of the detector. Color scale represents the magnitude of the superconducting current normalized by its maximum. Color white indicates where the value of the current is strictly zero.}
\label{fig_2}
\end{figure}

\begin{figure}[!t]
\centering
\includegraphics[width=2.5in]{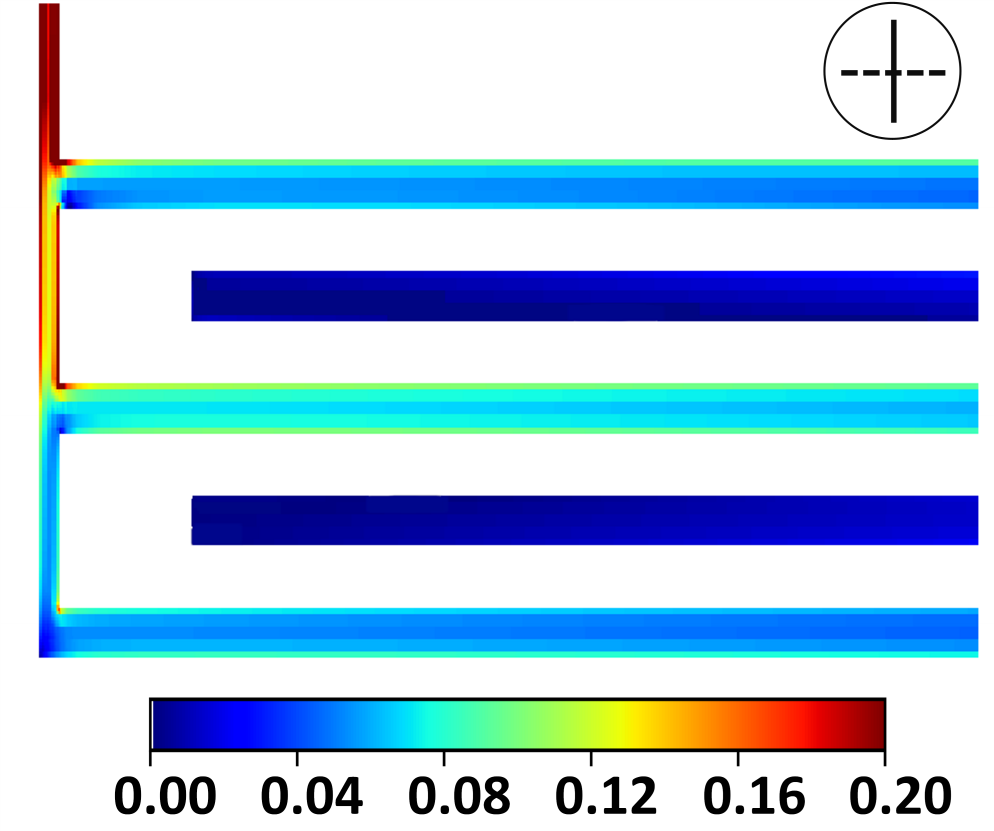}
\caption{Color plot showing the magnitude of the simulated currents in a zoomed region of the interdigitated capacitor. Color scale represents the magnitude of the current, which has been normalized to the maximum value in the inductor. The continuous line in the circular inset indicates the polarization direction aligned with the active region, while the dashed line indicates the cross-polarization direction, which is aligned with the interdigitated fingers in the capacitor.}
\label{fig_3}
\end{figure}

Electromagnetic simulations using Sonnet \cite{SONNET} were performed for both designs. Fig. \ref{fig_2}(a) depicts the amplitude of the simulated current distribution for the LER design. Fig. \ref{fig_2}(b) shows that the qLER maintains a near-uniform current distribution along the central meander, confirming that both detectors share an equivalent sensitive area. Fig. \ref{fig_3} depicts a zoomed area of the LER interdigitated capacitor, where parasitic currents are present. The simulation shows that the currents in the non-active area of the detector, the capacitor, are around fifteen times lower than the maximum value at the meander. Although they are lower in comparison, they may contribute to the response of the detector and hence, to reduce the polarization purity.

The qLER configuration offers significant advantages in terms of multiplexing capability and compact packaging. Unlike traditional lumped-element designs, which rely on large capacitive areas to tune the resonance frequency, the qLER's distributed resonator structure allows for more compact detectors. This is because the resonance frequency in distributed resonators depends primarily on a single geometric parameter (e.g., the meander length), enabling simpler frequency adjustment without the need for extensive capacitive modifications. As a result, the qLER design not only reduces pixel footprint but also enhances frequency allocation flexibility—an essential feature for large-format detector arrays. Furthermore, the quasi-lumped geometry improves fabrication scalability, making the qLER a promising configuration for high-density imaging and spectroscopic applications.

\section{Sample description}

\begin{figure}[!t]
\centering
\includegraphics[width=2.9in]{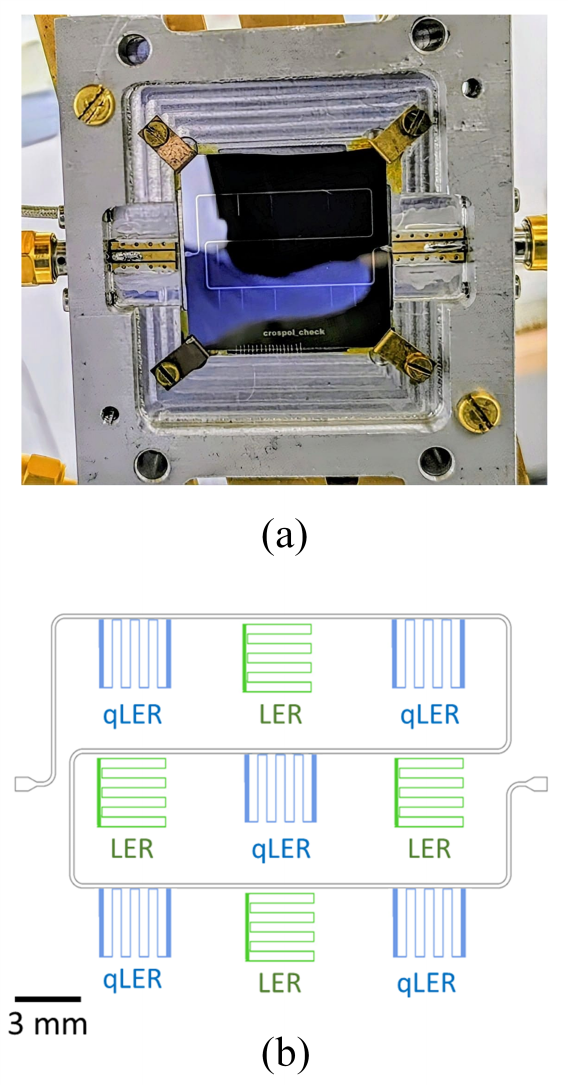}
\caption{(a) Layout of the two types of design. Green color represents the lumped detectors and blue color the quasi-lumped ones. Both types of detectors are interleaved in the chip and are placed with their longer meander sections perpendicular between them. (b) Optical image of the prototype mounted in the measurement holder prior to the cryogenic characterization.}
\label{fig_4}
\end{figure}

A 3×3 array demonstrator has been fabricated, integrating four KIDs with a LER-type design (LEKIDs) and five KIDs with a quasi-lumped half-wave resonator design (qLER). Both groups of detectors are coupled to a common readout microstrip transmission line. The complete design is presented in Fig. \ref{fig_4}(a). As explained in the previous section, the central absorber in the quasi-lumped detectors and in the lumped detectors share the same geometry, both matched to the free-space impedance in the W-band. Note that the two families are arranged with their long meander sections oriented perpendicularly, which enables their electrical discrimination through polarization response during cryogenic characterization.

The array was fabricated on a Ti/Al bilayer deposited on a 275 $\mu$m-thick silicon wafer, with a 200 nm aluminum backside layer serving as a reflective backshort and ground plane. The Ti layer is 10 nm thickness while Al is 15 nm. Further details on the nanofabrication process can be found in \cite{aja2020bilayer}. The superconducting bilayer shows a critical temperature of $T_c\sim$ 800 mK and a sheet resistance of $R_{s}$ = 0.9 $\Omega$/sq. This ensures sensitivity to frequencies above $\Delta_{TiAl} / h$ $\sim$ 55 GHz, making the device well-suited for W-band applications \cite{catalano2015bilayer}. The final developed device is shown in Fig. \ref{fig_4}(b) mounted prior to cryogenic characterization.

\begin{figure}[!b]
\centering
\includegraphics[width=0.9\columnwidth]{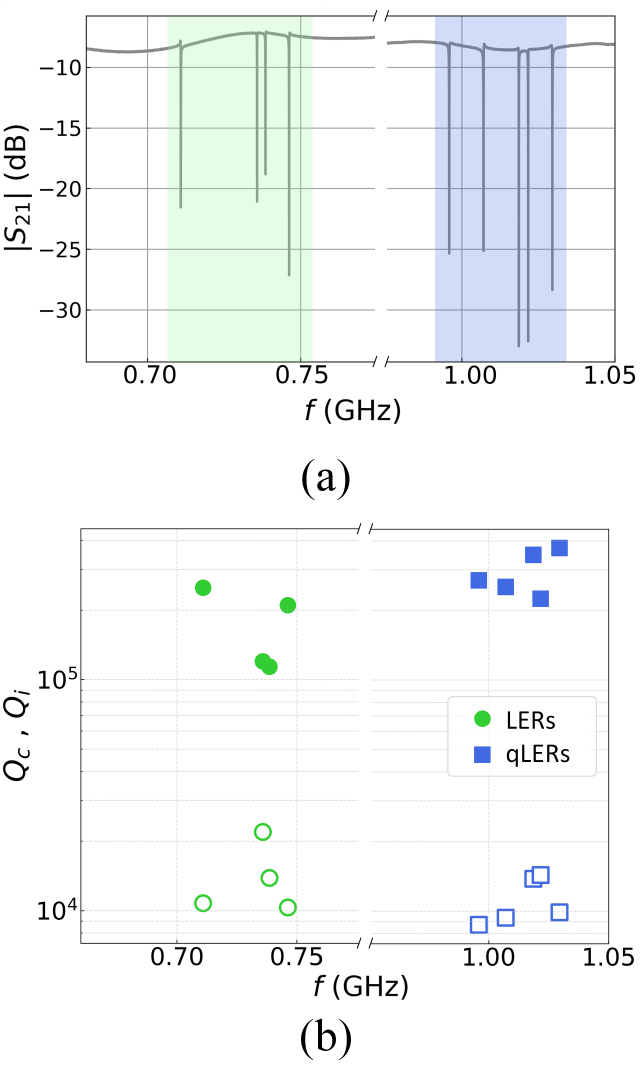}
\caption{(a) Amplitude of transmission as a function of frequency, showing resonances corresponding to both designs. LER resonances are highlighted in green at lower frequencies, while qLER resonances are indicated in blue. (b) Internal and external quality factors ($Q_i$, $Q_c$) for each detector family, using the same color scheme as in panel (a). $Q_i$ and $Q_c$ are represented with close and open symbols respectively.}
\label{fig_5}
\end{figure}

The resonance frequency of the detectors is defined as:

\begin{equation}
\label{deqn_ex1a}
f_0 = \frac{1}{2 \pi \sqrt{L C}}
\end{equation}

where $L$ is the inductance and $C$ the capacitance. The inductance of the resonator is a sum of the geometric inductance ($L_g$)—defined by design—and the kinetic inductance of the superconducting material ($L_k$). By comparing the resonances shown in Fig. \ref{fig_5}(a) with those obtained from the Sonnet simulations, the average value of $L_k$ is around 1.75 pH/sq. The $L_k$ value is obtained by comparing measurement and simulation results, more details on the procedure can be found in reference \cite{deOry2024low}.
The lumped approach offers direct control over $L$ and $C$, while in the quasi-lumped design the resonance frequency depends on the length of the planar resonator. Hence, for frequency multiplexing, different tuning strategies are used: for the LER type, the length of the last finger in the interdigitated capacitor is varied, changing the capacitance and shifting the resonance frequency; for the qLERs, frequency tuning is achieved by adjusting the total length of the resonator. The resonances of the detectors have been grouped around distinct readout frequency bands corresponding to the two detector types, enabling clear differentiation during characterization. Fig. 5(a) shows the transmission amplitude as a function of readout frequency across the spectral range containing all nine observable resonances measured at 100 mK. The transmission data was obtained using a Vector Network Analyzer (VNA) using a readout power signal with the resonance frequency of the characterized detector with a power of - 95 dBm at the chip level.

\begin{table}[!b]
\caption{Individual resonator parameters obtained from resonance analysis.\label{tab:table1}}
\centering
\begin{tabular}{c c c c c c c c c}
\hline
\hline
KID $\#$ & & $f_0 (MHz) $ & & $\kappa_i (Hz) $ && $Q_i$ & & $Q_c$\\
\hline
\hline
KID1 & & 710.929 & & 1423 & & 249770 & & 10758 \\
KID2 & & 735.832 & & 3064 & & 120069 & & 21907 \\
KID3 & & 738.567 & & 3247 & & 113718 & & 13855 \\
KID4 & & 746.296 & & 1777 & & 210042 & & 10322 \\
KID5 & & 995.865 & & 1842 & & 270345 & & 8718 \\
KID6 & & 1007.085 & & 1992 & & 252805 & & 9356 \\
KID7 & & 1018.577 & & 1463 & & 348126 & & 13792 \\
KID8 & & 1021.643 & & 2272 & & 224839 & & 14310 \\
KID9 & & 1029.533 & & 1381 & & 372826 & & 9867 \\
\hline
\hline
\end{tabular}
\color{black}
\end{table}

The parameters of the resonances, which characterize the resonator physics, have been assessed using the analysis method described in \cite{probst2015efficient}, by fitting the complex resonances with the following model:

\begin{equation}
\label{deqn_ex1a}
S_{21} = a e ^{i(2\pi f \tau + \tau_0)} \left[ 1 - \frac{\kappa_c e^{i\varphi}}{i(f-f_0) + \kappa} \right]
\end{equation}

where the first factor accounts for the attenuation ($a$) and the phase delay ($\tau$ and $\tau_0$) introduced by the experimental setup and the second factor accounts for the resonance itself. In this expression $\kappa_c$ represents the coupling rate to the transmission line and $\kappa_i$ the internal loss rate, being related to the coupling and internal quality factors following $Q_c = f_0 / 2\kappa_c$ and $Q_i = f_0 / 2\kappa_i$ respectively. The parameter $\varphi$ corresponds to the Fano phase \cite{rieger2023fano}, which represents the deviation from the ideal Lorentzian-shape of the resonance (i.e., the asymmetries in the resonance dip in the amplitude of the transmission). Table \ref{tab:table1} lists the resonance frequencies, quality factors and internal loss rates obtained for each detector using the analysis method described above. Fig. \ref{fig_5}(b) shows the extracted coupling values. The average coupling quality factor for all nine detectors is $Q_c \sim 12710$, which is in good agreement with the simulated ones. The lumped detectors exhibit an average internal quality factor of $Q_i \sim 1.6 \times 10^{5}$, while the quasi-lumped design reaches an average value of $Q_i \sim 2.7 \times 10^{5}$. It is worth noting that the increase of the internal loss rate for the qLER design arises from its higher resonance frequency, rather than from a decrease in the internal loss rate.

\section{Response characterization}

\begin{figure}[!t]
\centering
\includegraphics[width=\columnwidth]{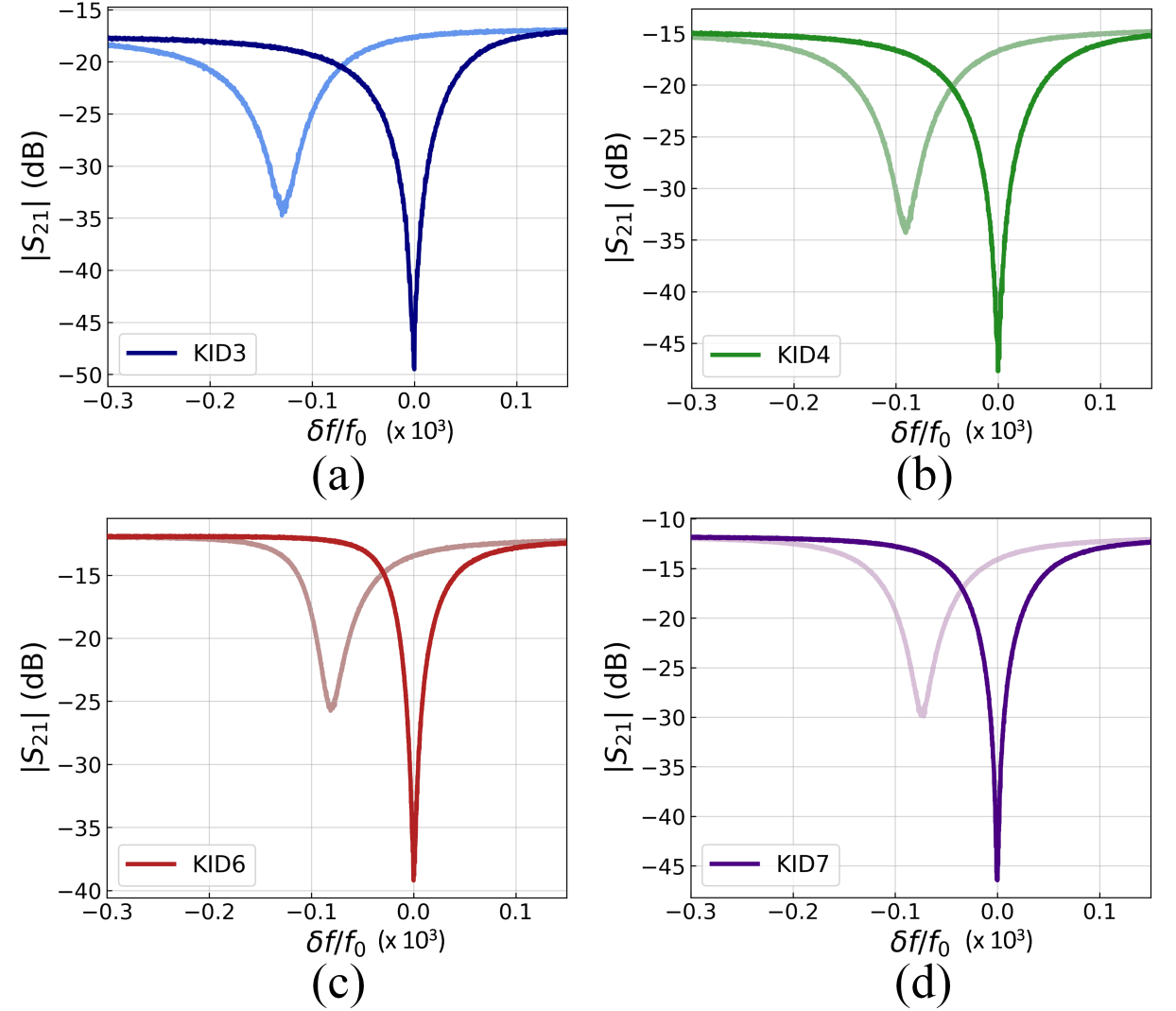}
\caption{Amplitude of the readout transmission as a function of frequency for LER-type detectors (a) KID3 and (b) KID4, and qLER-type detectors (c) KID6 and (d) KID7. Dark colors represent the response under low illumination load, while lighter colors represent the response under high load. Lighter color resonances appear shifted down towards negative values (i.e., to lower frequencies) due to the increase of the optical load. For each panel, horizontal axes have been normalized to the corresponding resonance frequency value at low optical loading.}
\label{fig_6}
\end{figure}

The optical response of both detector types was characterized by measuring their frequency response under varying illumination conditions using a variable-temperature blackbody source. The millimeter radiation reaches the bare absorbers, as in the case of filled-array cameras, where no focusing elements such as lenses or horns are employed. Fig. \ref{fig_6} compares the normalized frequency shift for both the LER and qLER detectors under changes in the optical load. The plots show the transmission amplitude as a function of the readout frequency for four representative detectors - two of each design - illuminated by a cold ($\sim$ 20 K)  and a warm ($\sim$ 300 K) black-body source. The observed frequency shifts reflect the change in the absorbed power as a result of varying radiation power, providing a direct measure of the responsivity of the detectors.

Despite the minor differences in the current distribution of the detector, both types show a similar response. This is the result of both designs sharing a similar sensitive area. The estimated difference in absorbed power ($\delta P_{abs}$) between low and high illumination loads is $\delta P_{abs} \sim 1.7$ pW, calculated using a ray-tracing software and considering the filters between the black-body radiation source and the device. This value gives an average responsivity ($R = \delta f / \delta P_{abs}$) of $R \sim 4.78\cdot10^{16} Hz/W$ for the lumped-type detectors and $R \sim 4.54\cdot10^{16} Hz/W$. These results confirm that the quasi-lumped design maintains the effective optical performance while offering advantages in layout compactness and multiplexability.

\section{Polarization discrimination}
\begin{figure}[!t]
\centering
\includegraphics[width=0.9\columnwidth]{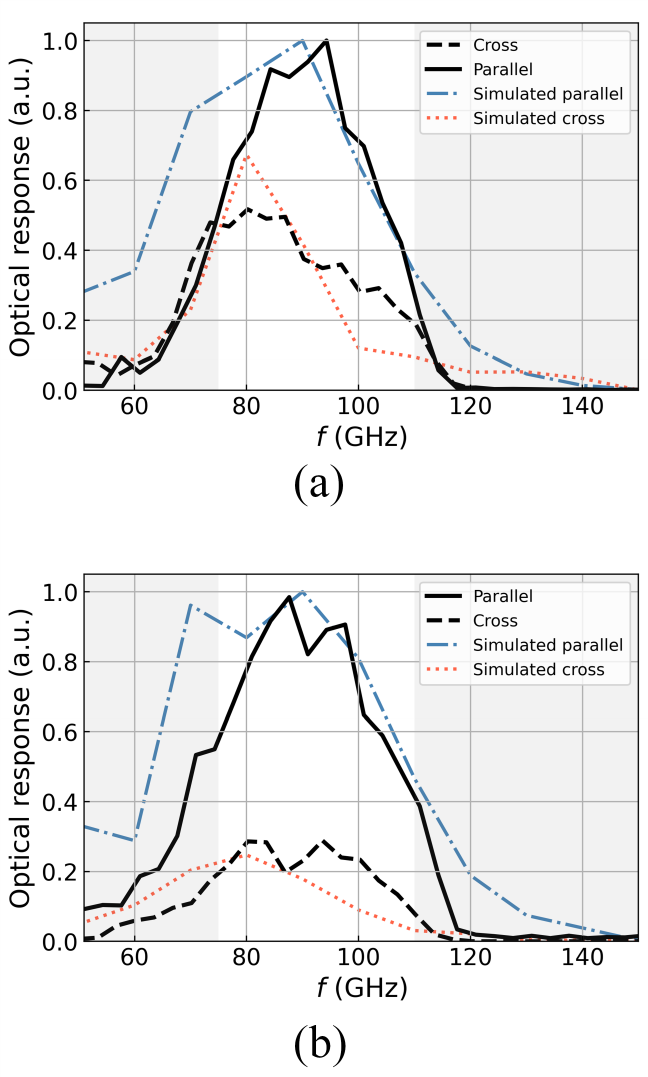}
\caption{Normalized spectral response for (a) LER-type and (b) qLER-type detectors, representing the KID frequency shift as a function of the incident optical frequency. Data represent the average response across all detectors of each type. Solid lines indicate the response for radiation polarized parallel to the meander’s long axis; dashed lines show the cross-polarization response. The white region represents the W-band (75 -110 GHz).  Dash-dot lines show the simulated spectral response for the parallel (blue) and cross (red) polarization. }
\label{fig_7}
\end{figure}

The polarimetric spectral response of both detector types was characterized under cryogenic conditions using a room-temperature Martin–Puplett interferometer (MPI). A detailed schematic and operational description of the MPI setup can be found in the supplementary material of Maleeva et al \cite{maleeva2018circuit}. Fig. \ref{fig_7} presents the normalized spectral response for the two detector families and their co- and cross- polarized responses. The curves shown represent the average response obtained by combining the spectra from all detectors of each type, providing a representative comparison of their behavior. The absortion under a plane-wave incidence has been simulated using CST with periodic boundary conditions. As depicted in Fig. \ref{fig_7}, the simulations are in good agreement with the measurements \sout{as shown in }.

The distributed detectors show an improved cross-polarization discrimination (XPD) compared with the non-distributed ones. The cross-polarization discrimination value has been defined as the ratio between the cross-polarized and co-polarized spectral responses both integrated along the W-band (white region in Fig. \ref{fig_7}). The average XPD value of the detectors comprising an interdigitated capacitor is -2.8 dB. With the qLER design, this value decreases down to -5.1 dB, proving that parasitic currents in the capacitor of the LER design contribute to cross-polarization detection.

\section{Conclusion}
This work is focused on understanding the origin of the cross-polarization effects in meandered LEKIDs and the possible contribution of the interdigital capacitor to the optical response. For this purpose, a comparative study of lumped-element and quasi-lumped KID—where no capacitor is employed—architectures is presented. Cryogenic and optical characterization confirms that both designs exhibit similar response behaviour due to their shared optical absorption geometry. However, the qLER design outperforms in terms of cross-polarization discrimination, achieving an average value of –5.1 dB compared to –2.8 dB in the LER. This work confirms the hypothesis that parasitic currents in the interdigitated capacitor lower the polarization fidelity, introducing an undesired cross-polarization response. While this result supports a design strategy based on removing capacitive elements from the pixel layout to reduce polarization leakage, the current XPD performance remains insufficient for polarization-sensitive instruments. A value of at least –20 dB is typically required for a high-fidelity polarimetric camera. Therefore, further improvements are still necessary. One possible direction is to replace the shorter segments of the meandered inductor, which are also known to introduce cross-polar poisoning, with a sawtooth structure \cite{deory2023}. The implementation of hybrid resonators \cite{janssen2013high}, in particular using another superconducting material with a larger energy gap in the structures that contribute to the polarization leakage, or implementing parallel plate capacitors \cite{beldi2019high}, could also help to improve the XPD performance. Alternatively, the addition of coupling systems like horns \cite{mccarrick2018design} or lenses \cite{llombart2015fourier} may help to focus radiation onto co-polarized absorbing structures, and avoid the exposure to radiation of cross-polarized ones, thus isolating their unwanted response.

\section*{Acknowledgments}
The authors would like to thank Ricardo Ferrándiz from Instituto Nacional de Técnica Aeroespacial for the chip holder design, fabrication and assembly. They would also thank the IMDEA-Nanoscience cleanroom for support in nanofabrication of the chip.

\printbibliography

%
%
%
%
%


\end{document}